\documentclass[prd,aps,amssymb,showpacs]{revtex4}

\newcommand{\beq}{\begin{equation}}
\newcommand{\eeq}{\end{equation}}
\newcommand{\bear}{\begin{eqnarray}}
\newcommand{\ear}{\end{eqnarray}}
\newcommand{\earn}{\nonumber \end{eqnarray}}

\newcommand{\gsim}{\mathop{\lefteqn{\raise.9pt\hbox{$>$}}
\raise-3.7pt\hbox{$\sim$}}}
\newcommand{\lsim}{\mathop{\lefteqn{\raise.9pt\hbox{$<$}}
\raise-3.7pt\hbox{$\sim$}}}
\arraycolsep=\mathsurround

\begin{document}

\title{Quantum-corrected ultraextremal horizons and validity of WKB in
massless limit}
\author{Arkady A. Popov}
\email{apopov@kzn.ru} \affiliation{Department of Mathematics,
Kazan State Pedagogical University, 1 Mezhlauk St., Kazan
420021, Russia}
\author{O. B. Zaslavskii}
\email{ozaslav@kharkov.ua} \affiliation{Astronomical Institute of
Kharkov V.N. Karazin National University, 35 Sumskaya St.,
Kharkov, 61022, Ukraine}

\begin{abstract}
We consider quantum backreaction of the quantized scalar field with an
arbitrary mass and curvature coupling on ultraextremal horizons. The problem
is distinguished in that (in contrast to non-extremal or extremal black
holes) the WKB approximation remains valid near $r_{+}$ (which is the radius
of the horizon) even in the massless limit. We examine the behavior of the
stress-energy tensor of the quantized field near $r_{+}$ and show that
quantum-corrected objects under discussion do exist. In the limit of the
large mass our results agree with previous ones known in literature.
\end{abstract}

\maketitle




\section{Introduction}

The distinguished role of extremal horizons (EH) is beyond any doubts. It is
sufficient to mention briefly such issues as black hole entropy, the
scenarios of evaporation including the nature of remnants, etc. Meanwhile,
although such object appear naturally on the pure classical level (the
famous examples is the Reissner-Nordstr\"{o}m black hole with the mass equal
to charge), the question of their existence becomes non-trivial in the
semiclassical case, when backreaction of quantum fields (whatever weak it
be) is taken into account. This is due to the fact that the
quantum-corrected metric contains some combinations of the stress-energy
tensor having the meaning of the energy measured by a free-falling observer
that potentially may diverge near the extremal horizon. However, numerical
calculations showed that such divergencies do not occur for massless fields
in the Reissner-Nordstr\"{o}m background \cite{andRN}. Analytical studies
for massive quantized fields \cite{MZ2} gave the same result. Recently they
have been extended to so called ultraextremal horizons (UEH) \cite{MZ1} when
the metric coefficient
\begin{equation}
-g_{tt} \sim (r_{+}-r)^{3}  \label{n}
\end{equation}
near the horizon (here $r$ is the Schwarzschild-like coordinate, $r=r_{+}$
corresponds to the horizon). Such horizons are encountered, for example, in
the Reissner-Nordstr\"{o}m-de Sitter solution, when the cosmological
constant $\Lambda >0$ \cite{Romans}. In doing so, it turns out that the
horizon is of cosmological nature, so $r$ approach $r_{+}$ from $r<r_{+}$.

The results for UEH are obtained in \cite{MZ1} for massive fields only. The
natural question arises, whether semiclassical UEH exist for quantum fields
of an arbitrary mass, including the massless case. It is worth reminding
that the mathematical basis for analytical calculations of the stress-energy
tensor of quantum fields consists in using WKB approximation. The well-known
DeWitt-Schwinger approximation \cite{AHS, Mat1, Mat2} is expansion with
respect to the small parameter
\begin{equation}
\epsilon =\left( \lambda /r_{0}\right) ^{2},
\end{equation}
where $\lambda $ is the Compton wavelength, $r_{0}$ is the characteristic
curvature scale of the gravitational field, near the horizon $r_{0}=r_{+}$.
For massive fields $\lambda \sim m^{-1}$, where $m$ is the mass of field (we
use the geometric system of units with $G=c=\hbar=1$). For massless fields $%
\lambda \sim r_{+}$, so $\epsilon $ ceases to be a small parameter. On the
first glance, the same reasons that prevent analytical studies of
semiclassical extremal black holes should reveal themselves here as well, so
if we are unable to cope with usual EH, the situation is not better with
UEH. However, the parameter WKB expansion can be redefined in such a way
that this parameter will be still small near UEH in contrast to usual EH.
This is the key observation that enables us to perform analytical
calculations and establish that semiclassical UEH do exist for an arbitrary
mass of quantized field.

\section{Validity of WKB approximation}

The metric under consideration reads
\begin{equation}
ds^{2}=-U(r)dt^{2}+\frac{dr^{2}}{V(r)}+r^{2}\left( d\theta ^{2}+\sin
^{2}\theta d\varphi ^{2}\right) .  \label{1a}
\end{equation}
We denote the root of equation $V(r)=0$ as $r_{+}$, i.e.
\begin{equation}
V(r_{+})=0.  \label{v=0}
\end{equation}
The applicability of the WKB approximation requires that the wavelength of
the particle must vary only slightly over distances of the order of itself
(see, e.g., Sec. 46 of \cite{ll}). It is convenient to formulate the
corresponding condition in terms of the invariant proper distance $\rho \,$.
If we assume that the corresponding scale on which the metric changes is
determined by first derivatives (this is the case in the problem under
consideration) this scale is characterized by the parameter
\begin{equation}
r_{0}^{-1}=\max \left\{ \left\vert \frac{d\ln (U)}{d\rho
}\right\vert ,\ \left\vert \frac{d\ln (r)}{d\rho }\right\vert
\right\}.  \label{Lm}
\end{equation}
For the scalar quantum field of arbitrary mass in static spherically
symmetric spacetimes we can use the WKB approximation derived in \cite%
{Popov, ps} (see, also, \cite{Khat1, Khat2} for the fields of spin 1 and
1/2). Then the role of the Compton wavelenght is played by the parameter%
\begin{equation}
\lambda =\left[ m^{2}+\frac{2\xi }{r^{2}}\right] ^{-1/2},  \label{lam}
\end{equation}
$\xi $ is the conformal coupling parameter. Let us note, that in the case
\begin{equation}
m^{2}\gg \frac{2\xi }{r^{2}}
\end{equation}
the approximations of \cite{Popov, ps} coincide with corresponding results
of \cite{AHS, Mat1, Mat2}.

Let in the zero approximation the metric have the form (\ref{1a}) with $U=V$%
. Then $r_{+}$ corresponds to the event horizon. We assume that
\begin{equation}
\text{ }\frac{d^{k}V}{dr^{k}}_{|r=r_{+}}=0\text{, }k=0,1,2,...n-1.
\label{V}
\end{equation}
Near the horizon $r\rightarrow r_{+}$ and the derivatives of $r^{2}$ and $V$
with respect to $\rho $ can be evaluated as follows
\begin{equation}
\left\vert {\frac{dr^{2}}{d\rho }}\right\vert \ \sim \ \frac{|r-r_{+}|^{n/2}%
}{r_{+}^{n/2-1}},  \label{val1}
\end{equation}
\begin{equation}
\left\vert \frac{1}{U}{\frac{dU}{d\rho }}\right\vert \ \sim \ \frac{%
|r-r_{+}|^{n/2-1}}{r_{+}^{n/2}}.  \label{val2}
\end{equation}
The case $n=2$ corresponds to the extremal Reissner-Nordstr\"{o}m metric.
Then, it follows from (\ref{val1}) that $\left\vert {\frac{dr^{2}}{d\rho }}%
\right\vert \rightarrow 0$ but $\left\vert \frac{1}{U}{\frac{dU}{d\rho }}%
\right\vert \sim r_{+}^{-1}$ does not contain the small parameter. As a
result, we obtain from (\ref{Lm}) that $r_{0}\sim r_{+}$. Therefore, for
massless fields, when $\lambda \sim r_{+}$, the parameter $\varepsilon \sim
1 $ ceases to be small and the WKB approximation fails.

However, the situation changes drastically, if $n>2$. Then
\begin{equation}
\varepsilon =\frac{\lambda ^{2}}{r_{+}^{2}}\left( \frac{r-r_{+}}{r_{+}}%
\right) ^{n-2}\ll 1.  \label{e}
\end{equation}

The approximation becomes increasingly accurate as the ratio of the $\lambda
^{2}/r_{0}^{2}$ approaches zero, i.e., in the limit $r\rightarrow r_{+}$.
Thus, the WKB approximation is valid near $r_{+}$ for any finite mass $m$ of
the quantized field, including $m=0$. Below we use this approach for finding
the quantum-corrected metric.


\section{Quantum backreaction near ultraextremal horizon}


The 00 and 11 Einstein equation for our system read%
\begin{equation}
\frac{V^{\prime }}{r}+\frac{V}{r^{2}}-\frac{1}{r^{2}}+\Lambda =8\pi
T_{t}^{t},  \label{one}
\end{equation}
\begin{equation}
\frac{VU^{\prime }}{rU}+\frac{V}{r^{2}}-\frac{1}{r^{2}}+\Lambda =8\pi
T_{r}^{r},  \label{two}
\end{equation}
where prime denotes the derivative with respect to $r$, $T_{\mu }^{\nu
}=T_{\nu }^{\mu \,(cl)}+\langle T_{\nu }^{\mu }\rangle _{ren}$, where $%
T_{\nu }^{\mu \,(cl)}$ is the classical source, $\langle T_{\nu }^{\mu
}\rangle _{ren}$ is the renormalized stress-energy tensor of quantized
fields.

We will be dealing with the electromagnetic field, then
\begin{equation}  \label{cl}
T_{\nu }^{\mu \,(cl)}=\frac{Q^{2}}{8 \pi r^{4}}diag(-1,-1,1,1).
\end{equation}

We assume the conditions
\begin{equation}
U(r_{+})=U^{\prime }(r_{+})=U^{\prime \prime }(r_{+})=V(r_{+})=V^{\prime
}(r_{+})=V^{\prime \prime }(r_{+})=0  \label{ultra}
\end{equation}
which correspond to the UEH with $n=3$. This choice is motivated by its
physical interpretation since it corresponds to the ultraextremal
(ultracold) horizon of the Reissner-Nordstr\"{o}m-de Sitter metric \cite%
{Romans}.

If backreaction is neglected,
\begin{equation}
U(r)=V(r)= -\frac{\left( r+3r_{+}\right) }{6{r}_{+}\!{}^{2}\,r^{2}}\left(
r-r_{+}\right) ^{3}  \label{3a}
\end{equation}
where the unperturbed conditions of ultraextremalization read \cite{Romans}
\begin{equation}
Q^{2}=\frac{{r}_{+}\!{}^{2}}{2}\,,  \label{ql}
\end{equation}
\begin{equation}
\Lambda =\frac{1}{2 {r}_{+}\!{}^{2}}.  \label{lr}
\end{equation}

To evaluate the role of backreaction, we proceed along the same lines as in
\cite{MZ2,MZ1}, i.e we start not from a classical background with further
adding quantum corrections but from the quantum-corrected self-consistent
geometries from the very beginning. It means that for $\varepsilon \neq 0$,
we still have the UEH although the explicit conditions of
ultraextremalization differ from (\ref{ql}), (\ref{lr}) by the terms of the
order $\varepsilon $.

The solutions of equation (\ref{one}) can be written as follows:
\begin{equation}  \label{W1}
V(r)=1+\left( \frac{\Lambda r_{+}^{3}}{3}-\frac{Q^{2}}{r_{+}}-r_{+}\right)
\frac{1}{r}-\frac{\Lambda r^{2}}{3}+\frac{Q^{2}}{r^{2}}+\frac{8\pi }{r}%
\int_{r_{+}}^{r}d\tilde{r}\tilde{r}^{2}\langle T_{t}^{t}\rangle _{ren},
\label{V1}
\end{equation}
where $\langle T_{\nu }^{\mu }\rangle _{ren}$ is calculated on the
unperturbed background and it follows from the conditions of
ultraextremalization (\ref{ultra}) that
\begin{equation}
Q^{2}=\frac{r_{+}^{2}}{2}-2\pi r_{+}^{5}\frac{d\langle T_{t}^{t}\rangle
_{ren}}{dr}_{|r=r_{+}},
\end{equation}
\begin{equation}
\Lambda =\frac{1}{2r_{+}^{2}}+8\pi {\langle T_{t}^{t}\rangle _{ren}}%
_{|r=r_{+}}+2\pi r_{+}\frac{d\langle T_{t}^{t}\rangle _{ren}}{dr}_{|r=r_{+}}.
\end{equation}
The solutions of equation (\ref{two}) can be written as follows:
\begin{equation}
U=\exp (2\psi )V \approx (1+2 \psi) V,  \label{uv}
\end{equation}
\begin{equation}  \label{F1}
\psi =\frac{const}{2}+4\pi \int_{r_+}^{r} d \tilde r F(\tilde r)\text{, }%
F(\tilde r) =\tilde r \frac{T_{r}^{r}-T_{t}^{t}}{V}.
\end{equation}
Then it follows from (\ref{cl}) that non-vanishing contribution in $F(r)$
comes from quantum fields only. Below we restrict ourselves by the case of
the scalar field.

Now the crucial question is the behavior of $F(r)$ near the ultraextremal
horizon. If it remains finite at $r_{+}$, the metric function $U(r)$ (\ref%
{uv}) has the same asymptotic as $V(r)$ that corresponds to the
ultraextremal horizon. As far as the choice of the state is concerned, it is
worth reminding (see Sec. 11.3.7 of \cite{nf}) that the DeWitt-Schwinger
approximation for the very massive field \cite{AHS, Mat1, Mat2} in a given
background is almost state-independent and entirely local, depending at each
point only on the values of the curvature and its derivatives. However, the
reservation "almost" is crucial now since the difference between
Hartle-Hawking, Boulware and Unruh states is essential in the very vicinity
of the horizon. Meanwhile, as we are intending to elucidate whether the
ultraextremal horizon exists or not, it is just this vicinity that is
crucial for our purposes. Apart from this, we discuss in general a finite
mass of quantum fields. Therefore, we would like to stress that we are
interested in the Hartle-Hawking state. The latter implies that we consider
the static region of the spacetime. Now it is confined by $0<r\leq r_{+}$.
The problem connected with the presence of singularity ar $r=0$ can be solve
in the same way as in \cite{MZ1}: we smear or simply replace it by some
central body with a regular center and the boundary at $r=R<r_{+}$. Then for
$R<r\leq r_{+}$ one can consider propagation of quantum fields in the
everywhere regular background. As, is explained above, the WKB approximation
does work now, we can use safely the formulas for the renormalized
expression for $\langle T_{\nu }^{\mu }\rangle _{ren}$ of the quantized
scalar field obtained in \cite{Popov} for any finite $m$ and arbitrary
coupling $\xi $ to the scalar curvature, applying them to the near-horizon
region
\begin{equation}
\varepsilon =\frac{\lambda ^{2}}{r_{+}^{2}}\left( \frac{r-r_{+}}{r_{+}}%
\right) \ll 1.
\end{equation}

It turns out that in the expansion%
\begin{equation}
{\langle }T_{\nu }^{\mu }\rangle _{ren}={\langle }T_{\nu }^{\mu }\rangle
_{ren\mid r=r_{+}}+\frac{d\langle T_{\nu }^{\mu }\rangle _{ren}}{dr}%
_{|r=r_{+}}(r-r_{+})+\frac{1}{2}\frac{d^{2}\langle T_{\nu }^{\mu }\rangle
_{ren}}{dr^{2}}_{|r=r_{+}}(r-r_{+})^{2}+...  \label{ta}
\end{equation}
the coefficients at $(r-r_{+})^{k}$ with $k=0,1,2$ coincide for ${\langle }%
T_{t}^{t}\rangle _{ren}$ and ${\langle }T_{r}^{r}\rangle _{ren}$. As a
result, the difference $\langle T_{r}^{r}\rangle _{ren}-\langle
T_{t}^{t}\rangle _{ren}=B(r_{+}-r)^{3}$ $+O((r_{+}-r)^{4})$ with the finite
constant $B$ has the same order as $V$, so that $F$ turns out to be finite
as $r\rightarrow r_{+}$. (See technical details below.) Thus, the quantity $%
\psi $ is also finite, so that the quantum-corrected UEH do exist.

One can find also the explicit expression for the quantum-corrected metric
in terms of $r_{+}$ and $\langle T_{\nu }^{\mu }\rangle _{ren}$ near $r_{+}$%
. Integration of (\ref{W1}, \ref{uv}, \ref{F1}) gives us
\begin{eqnarray}
V(r) &=&\left[ -\frac{2}{3r_{+}^{3}}+\frac{20\pi }{3}\frac{d\langle
T_{t}^{t}\rangle _{ren}}{dr}_{|r=r_{+}}+\frac{4\pi r_{+}}{3}\frac{%
d^{2}\langle T_{t}^{t}\rangle _{ren}}{dr^{2}}_{|r=r_{+}}\right] (r-r_{+})^{3}
\label{Ve} \\
&&+\left[ \frac{7}{6r_{+}^{4}}-\frac{20\pi }{3r_{+}}\frac{d\langle
T_{t}^{t}\rangle _{ren}}{dr}_{|r=r_{+}}+\frac{2\pi }{3}\frac{d^{2}\langle
T_{t}^{t}\rangle _{ren}}{dr^{2}}_{|r=r_{+}}\right.   \nonumber \\
&&\left. +\frac{\pi r_{+}}{3}\frac{d^{3}\langle T_{t}^{t}\rangle _{ren}}{%
dr^{3}}_{|r=r_{+}}\right] (r-r_{+})^{4}+O\left( \frac{(r-r_{+})^{5}}{%
r_{+}^{5}}\right) ,
\end{eqnarray}%
\begin{eqnarray}
U(r) &=&\left[ -\frac{2(1+const)}{3r_{+}^{3}}+\frac{20\pi }{3}\frac{d\langle
T_{t}^{t}\rangle _{ren}}{dr}_{|r=r_{+}}+\frac{4r_{+}\pi }{3}\frac{%
d^{2}\langle T_{t}^{t}\rangle _{ren}}{dr^{2}}_{|r=r_{+}}\right] (r-r_{+})^{3}
\nonumber  \label{Ue} \\
&&+\left[ \frac{7(1+const)}{6r_{+}^{4}}-\frac{20\pi }{3r_{+}}\frac{d\langle
T_{t}^{t}\rangle _{ren}}{dr}_{|r=r_{+}}+\frac{2\pi }{3}\frac{d^{2}\langle
T_{t}^{t}\rangle _{ren}}{dr^{2}}_{|r=r_{+}}\right.   \nonumber \\
&&\left. +\frac{\pi r_{+}}{3}\frac{d^{3}\langle T_{t}^{t}\rangle _{ren}}{%
dr^{3}}_{|r=r_{+}}+\frac{4\pi r_{+}}{3}\left( \frac{d^{3}\langle
T_{r}^{r}\rangle _{ren}}{dr^{3}}_{|r=r_{+}}-\frac{d^{3}\langle
T_{t}^{t}\rangle _{ren}}{dr^{3}}_{|r=r_{+}}\right) \right] (r-r_{+})^{4}
\nonumber \\
&&+O\left( \frac{(r-r_{+})^{5}}{r_{+}^{5}}\right).
\end{eqnarray}

Now we may substitute into these general formulas for the quantum-corrected
metric explicit expressions for the stress-energy tensor from Appendix. As,
in general, the corresponding expressions are very cumbersome, we restrict
ourselves by two particular case.

\subsection{Massless fields with conformal coupling $\left(\xi =1/6
\right)$}

The expressions (\ref{Ve}), (\ref{Ue}) read now
\begin{eqnarray}
V(r) &=&\left[ -\frac{2}{3r_{+}^{3}}-\frac{0.00372}{\pi r_{+}^{5}}\right]
(r-r_{+})^{3}+\left\{ \frac{7}{6r_{+}^{4}}\right.   \nonumber  \label{Vm0} \\
&&\left. -\frac{20}{3\pi r_{+}^{6}}\left[ 0.03058-\frac{1}{120}\ln \left( m_{%
\mbox{\tiny \sl DS}}^{2}r_{+}^{2}\right) \right] \right\}
(r-r_{+})^{4}+O\left( \frac{(r-r_{+})^{5}}{r_{+}^{5}}\right) ,
\end{eqnarray}%
\begin{eqnarray}
U(r) &=&\left[ -\frac{2(1+const)}{3r_{+}^{3}}-\frac{0.00372}{\pi r_{+}^{5}}%
\right] (r-r_{+})^{3}+\left\{ \frac{7(1+const)}{6r_{+}^{4}}\right.
\nonumber  \label{Um0} \\
&&\left. +\frac{1}{\pi r_{+}^{6}}\left[ -0.26311+\frac{7}{90}\ln \left( m_{%
\mbox{\tiny \sl DS}}^{2}r_{+}^{2}\right) \right] \right\}
(r-r_{+})^{4}+O\left( \frac{(r-r_{+})^{5}}{r_{+}^{5}}\right) .
\end{eqnarray}

\subsection{ Massive fields $ \left( m_{\mbox{\tiny \sl DS}}^{2}=m^{2}\gg 2
\xi /r_{+}^{2} \right)$ }

The quantum-corrected expressions (\ref{Ve}), (\ref{Ue}) for the massive
field case are now
\begin{eqnarray}
V(r) &=&\left[ -\frac{2}{3r_{+}^{3}}+\frac{1}{m^{2}\pi r_{+}^{7}}\left( -%
\frac{\xi }{135}+\frac{1}{378}\right) \right] (r-r_{+})^{3}  \nonumber \\
&&+\left[ \frac{7}{6r_{+}^{4}}+\frac{1}{m^{2}\pi r_{+}^{8}}\left( \frac{\xi
}{60}-\frac{1}{360}\right) \right] (r-r_{+})^{4}+O\left( \frac{(r-r_{+})^{5}%
}{r_{+}^{5}}\right),
\end{eqnarray}%
\begin{eqnarray}
U(r) &=&\left[ -\frac{2(1+const)}{3r_{+}^{3}}+\frac{1}{m^{2}\pi r_{+}^{7}}%
\left( -\frac{\xi }{135}+\frac{1}{378}\right) \right] (r-r_{+})^{3}
\nonumber \\
&&+\left[ \frac{7(1+const)}{6r_{+}^{4}}+\frac{1}{m^{2}\pi
r_{+}^{8}}\left( \frac{5\xi }{108}-\frac{149}{7560}\right) \right]
(r-r_{+})^{4}+O\left( \frac{(r-r_{+})^{5}}{r_{+}^{5}}\right).
\end{eqnarray}

\section{Conclusion}

Usually, the validity of the WKB approximation and treatment of massless
quantized field (or fields with a finite mass) conflict with each other in
the region of strong gravitation field, in particular near the event
horizon. Nonetheless, we showed that, happily, there exist exceptions of
physical interest when both issues are reconciled. Using the expressions for
the stress-energy tensor found on the basis of the WKB approximation, we
showed that semiclassical (quantum-corrected) ultraextremal horizons exist
for any mass of the field and for any power $n>2$ in the asymptotic
expansion of the metric coefficient (\ref{n}) near the horizon. In doing so,
one can take $n=2+\delta $ where $\delta $ is as small as one likes. As
there is no doubt that semiclassical non-extremal black holes exist, it
turns out that both for $n<2$ and $n>2$ the horizon is well-defined. We
consider this as a strong (although not quite rigorous) argument in favour
of the existence of semiclassical extremal ($n=2$) black holes dressed by
quantum fields with a finite or even zero mass, in addition to numeric
results found in \cite{andRN}.

\section{Appendix. Explicit behavior of stress-energy tensor near UEH}

Here we list explicitly the coefficients in the expansion of relevant
components of $\langle T_{\nu }^{\mu }\rangle _{ren}$ near UEH$.$ They are
obtained from \cite{Popov} where calculations have been done under the
condition%
\begin{equation}
\mu _{+}^{2}=m^{2}r_{+}^{2}+2\xi -1/4>0.  \label{cond}
\end{equation}
We have:
\begin{eqnarray}
\langle T_{t}^{t}\rangle _{ren} &=&\frac{1}{\pi ^{2}r_{+}^{4}}\left\{ \left(
\xi -\frac{1}{8}\right) \frac{m^{2}r_{+}^{2}}{32}+\frac{3\xi ^{2}}{32}-\frac{
11\xi }{384}+\frac{79}{30720}+\left[ -\frac{m^{4}r_{+}^{4}}{64}\right.
\right.   \nonumber  \label{1tt} \\
&&\left. -\left( \xi -\frac{1}{6}\right) \frac{m^{2}r_{+}^{2}}{16}-\frac{\xi
^{2}}{16}+\frac{\xi }{48}-\frac{1}{480}\right] \ln \left( \frac{\mu _{+}^{2}
}{m_{\mbox{\tiny \sl DS}}^{2}r_{+}^{2}}\right) \left. +\frac{\mu _{+}^{4}}{8}
\left[ I_{1}(\mu _{+})-I_{2}(\mu _{+})\right] \right\}   \nonumber \\
&&+\frac{1}{\pi ^{2}r_{+}^{5}}\left\{ \frac{1}{\mu _{+}^{2}}\left[ \left( -
\frac{5\xi ^{2}}{24}+\frac{73\xi }{1152}-\frac{77}{23040}\right)
m^{2}r_{+}^{2}-\frac{5\xi ^{3}}{12}+\frac{103\xi ^{2}}{576}\right. \right.
\nonumber \\
&&\left. -\frac{53\xi }{2560}+\frac{7}{11520}\right] +\left[ \left( \xi -
\frac{1}{6}\right) \frac{m^{2}r_{+}^{2}}{8}+\frac{\xi ^{2}}{4}-\frac{\xi }{12
}+\frac{1}{120}\right] \ln \left( \frac{\mu _{+}^{2}}{m_{\mbox{\tiny \sl DS}
}^{2}r_{+}^{2}}\right)   \nonumber \\
&&+\mu _{+}^{2}\left[ \left( \xi -\frac{1}{8}\right) I_{2}(\mu _{+})+\left(
2\xi ^{2}-\frac{5\xi }{12}+\frac{1}{24}\right) I_{1}(\mu _{+})\right]
\nonumber \\
&&+\mu _{+}^{3}\left[ \left( \frac{m^{2}r_{+}^{2}}{8}+\xi ^{2}+\frac{65\xi }{
48}-\frac{59}{192}\right) \frac{dI_{1}(\mu _{+})}{d\mu _{+}}+\left( \frac{25
}{192}-\frac{31\xi }{48}-\xi ^{2}\right) \frac{dI_{0}(\mu _{+})}{d\mu _{+}}
\right.   \nonumber \\
&&\left. -\frac{m^{2}r_{+}^{2}}{8}\frac{dI_{2}(\mu _{+})}{d\mu _{+}}\right]
+\left( \xi -\frac{1}{4}\right) \mu _{+}^{4}\left[ \frac{23}{48}\frac{
d^{2}I_{1}(\mu _{+})}{d\mu _{+}^{2}}-\frac{5}{12}\frac{d^{2}I_{0}(\mu _{+})}{
d\mu _{+}^{2}}\right]   \nonumber \\
&&\left. +\left( \xi -\frac{1}{4}\right) \frac{\mu _{+}^{5}}{32}\left[ \frac{
d^{3}I_{1}(\mu _{+})}{d\mu _{+}^{3}}-\frac{d^{3}I_{0}(\mu _{+})}{d\mu
_{+}^{3}}\right] \right\} (r-r_{+})  \nonumber \\
&&+\frac{1}{\pi ^{2}r_{+}^{6}}\left\{ \left[ \frac{m^{4}r_{+}^{4}}{8}+\left(
\xi -\frac{1}{6}\right) \frac{m^{2}r_{+}^{2}}{4}\right] \mu _{+}\frac{
dI_{0}(\mu _{+})}{d\mu _{+}}-\frac{5}{4}\mu _{+}^{4}I_{2}(\mu _{+})\right.
\nonumber \\
&&+\left[ \frac{3m^{2}r_{+}^{2}}{8}+\frac{5}{2}\left( \xi -\frac{1}{8}
\right) \right] \mu _{+}^{2}I_{1}(\mu _{+})+\left[ -\left( \xi -\frac{1}{6}
\right) \frac{5m^{2}r_{+}^{2}}{16}-\frac{5\xi ^{2}}{8}+\frac{5\xi }{24}
\right.   \nonumber \\
&&\left. \left. -\frac{1}{48}\right] \ln \left( \frac{\mu _{+}^{2}}{m_{
\mbox{\tiny \sl DS}}^{2}r_{+}^{2}}\right) -\frac{m^{2}r_{+}^{2}}{192}+\frac{
5\xi ^{2}}{8}-\frac{35\xi }{192}+\frac{641}{46080}\right\} (r-r_{+})^{2}
\nonumber \\
&&+\frac{1}{\pi ^{2}r_{+}^{7}}\left\{ \left[ \frac{m^{6}r_{+}^{6}}{24}
+\left( \frac{5\xi }{12}-\frac{11}{144}\right) m^{4}r_{+}^{4}+\left( \frac{
7\xi ^{2}}{6}-\frac{13\xi }{36}+\frac{1}{30}\right) m^{2}r_{+}^{2}+\xi
^{3}\right. \right.   \nonumber \\
&&\left. -\frac{5\xi ^{2}}{12}+\frac{\xi }{15}-\frac{1}{168}\right] \frac{
d^{2}I_{0}(\mu _{+})}{d\mu _{+}^{2}}+\frac{1}{\mu _{+}}\left[ -\frac{
m^{6}r_{+}^{6}}{3}+\left( \frac{53}{288}-\frac{4\xi }{3}\right)
m^{4}r_{+}^{4}\right.   \nonumber \\
&&\left. +\left( -\frac{4\xi ^{2}}{3}+\frac{7\xi }{18}-\frac{11}{360}\right)
m^{2}r_{+}^{2}+\frac{\xi ^{2}}{24}-\frac{7\xi }{360}-\frac{1}{2520}\right]
\frac{dI_{0}(\mu _{+})}{d\mu _{+}}+\frac{5\mu _{+}^{4}}{2}I_{2}(\mu _{+})
\nonumber \\
&&+\left( -\frac{m^{2}r_{+}^{2}}{2}-\frac{9\xi }{2}+\frac{13}{24}\right) \mu
_{+}^{2}I_{1}(\mu _{+})+\left[ \left( \xi -\frac{1}{6}\right) \frac{
5m^{2}r_{+}^{2}}{8}+\frac{5\xi ^{2}}{4}-\frac{5\xi }{12}\right.   \nonumber
\\
&&\left. +\frac{11}{240}\right] \ln \left( \frac{\mu _{+}^{2}}{m_{
\mbox{\tiny
\sl DS}}^{2}r_{+}^{2}}\right) +\frac{1}{\mu _{+}^{2}}\left[ \frac{
11m^{4}r_{+}^{4}}{576}+\left( -\frac{5\xi ^{2}}{4}+\frac{125\xi }{288}-\frac{
2513}{69120}\right) m^{2}r_{+}^{2}\right.   \nonumber \\
&&\left. \left. -\frac{5\xi ^{3}}{2}+\frac{53\xi ^{2}}{48}-\frac{5483\xi }{
34560}+\frac{18653}{1935360}\right] \right\} (r-r_{+})^{3}+O\left(
(r-r_{+})^{4}\right) ,
\end{eqnarray}%
\begin{eqnarray}
\langle T_{r}^{r}\rangle _{ren} &-&\langle T_{t}^{t}\rangle _{ren}=\frac{1}{
\pi ^{2}r_{+}^{7}}\left\{ \frac{1}{\mu _{+}^{2}}\left[ \left( \frac{1}{144}-
\frac{5\xi }{144}\right) m^{2}r_{+}^{2}-\frac{5\xi ^{2}}{72}+\frac{11\xi }{
576}-\frac{37}{13440}\right] \right.   \nonumber  \label{00-11} \\
&&-\frac{1}{360}\ln \left( \frac{\mu _{+}^{2}}{m_{\mbox{\tiny \sl DS}
}^{2}r_{+}^{2}}\right) -\left( \xi -\frac{1}{6}\right) \mu _{+}^{2}I_{1}(\mu
_{+})+\left[ \left( \xi -\frac{1}{8}\right) \frac{m^{4}r_{+}^{4}}{3}\right.
\nonumber \\
&&\left. +\left( \frac{4\xi ^{2}}{3}-\frac{29\xi }{72}+\frac{23}{720}\right)
m^{2}r_{+}^{2}+\frac{4\xi ^{3}}{3}-\frac{23\xi ^{2}}{36}+\frac{19\xi }{180}-
\frac{11}{2520}\right] \frac{dI_{0}(\mu _{+})}{\mu _{+}d\mu _{+}}  \nonumber
\\
&&+\left[ -\left( \xi -\frac{1}{4}\right) \frac{m^{4}r_{+}^{4}}{6}+\left( -
\frac{2\xi ^{2}}{3}+\frac{2\xi }{9}-\frac{1}{45}\right) m^{2}r_{+}^{2}-\frac{
2\xi ^{3}}{3}+\frac{5\xi ^{2}}{18}-\frac{2\xi }{45}\right.   \nonumber \\
&&\left. \left. +\frac{1}{252}\right] \frac{d^{2}I_{0}(\mu _{+})}{d\mu
_{+}^{2}}\right\} \left( r-r_{+}\right) ^{3}+O\left( (r-r_{+})^{4}\right) .
\end{eqnarray}
It is essential that this difference has the same main order as the function
$V$, so that $F$ and $\psi $ in (\ref{F1}) are indeed finite on the horizon.
\begin{eqnarray}
\langle T_{\theta }^{\theta }\rangle _{ren} &=&\langle T_{\varphi }^{\varphi
}\rangle _{ren}=\frac{1}{\pi ^{2}r_{+}^{4}}\left\{ \frac{m^{2}r_{+}^{2}}{32}
\left( \xi -\frac{1}{8}\right) -\frac{1}{32}\left( \xi -\frac{1}{8}\right)
^{2}+\left[ -\frac{m^{4}r_{+}^{4}}{64}+\frac{\xi ^{2}}{16}\right. \right.
\nonumber  \label{eps} \\
&&\left. \left. -\frac{\xi }{48}+\frac{1}{480}\right] \ln \left( \frac{\mu
_{+}^{2}}{m_{\mbox{\tiny \sl DS}}^{2}r_{+}^{2}}\right) -\left( \xi -\frac{1}{
8}\right) \frac{\mu _{+}^{2}}{4}I_{1}(\mu _{+})+\frac{\mu _{+}^{4}}{8}
I_{2}(\mu _{+})\right\}   \nonumber \\
&&+\frac{1}{\pi ^{2}r_{+}^{5}}\left\{ \frac{1}{\mu _{+}^{2}}\left[ \left(
\frac{7\xi ^{2}}{24}-\frac{113\xi }{1152}+\frac{41}{4608}\right)
m^{2}r_{+}^{2}+\frac{7\xi ^{3}}{12}-\frac{143\xi ^{2}}{576}\right. \right.
\nonumber \\
&&\left. +\frac{271\xi }{7680}-\frac{77}{46080}\right] +\left[ \left( \frac{1
}{6}-\xi \right) \frac{m^{2}r_{+}^{2}}{8}-\frac{\xi ^{2}}{4}+\frac{\xi }{12}-
\frac{1}{120}\right] \ln \left( \frac{\mu _{+}^{2}}{m_{\mbox{\tiny \sl DS}
}^{2}r_{+}^{2}}\right)   \nonumber \\
&&+\left[ \left( 2\xi ^{2}+\frac{5\xi }{6}-\frac{3}{16}\right)
m^{2}r_{+}^{2}+4\xi ^{3}+\frac{13\xi ^{2}}{6}-\frac{5\xi }{6}+\frac{1}{16}
\right] I_{1}(\mu _{+})  \nonumber \\
&&-\left( \xi -\frac{1}{8}\right) \mu _{+}^{2}I_{2}(\mu _{+})+\left[ \left(
\xi ^{2}+\frac{47\xi }{48}-\frac{25}{128}\right) m^{2}r_{+}^{2}+2\xi ^{3}+
\frac{53\xi ^{2}}{24}\right.   \nonumber \\
&&\left. -\frac{73\xi }{96}+\frac{29}{512}\right] \mu _{+}\frac{dI_{1}(\mu
_{+})}{d\mu _{+}}+\left[ \left( -\xi ^{2}-\frac{13\xi }{48}+\frac{29}{384}
\right) m^{2}r_{+}^{2}-2\xi ^{3}\right.   \nonumber \\
&&\left. -\frac{19\xi ^{2}}{24}+\frac{35\xi }{96}-\frac{15}{512}\right] \mu
_{+}\frac{dI_{0}(\mu _{+})}{d\mu _{+}}+\frac{m^{2}r_{+}^{2}}{8}\mu _{+}^{3}
\frac{dI_{2}(\mu _{+})}{d\mu _{+}}  \nonumber \\
&&+\left( \frac{23\xi }{48}-\frac{7}{128}\right) \mu _{+}^{4}\frac{
d^{2}I_{1}(\mu _{+})}{d\mu _{+}^{2}}-\left( \frac{5\xi }{12}-\frac{7}{128}
\right) \mu _{+}^{4}\frac{d^{2}I_{0}(\mu _{+})}{d\mu _{+}^{2}}  \nonumber \\
&&\left. +\frac{\xi }{32}\mu _{+}^{5}\left[ \frac{d^{3}I_{1}(\mu _{+})}{d\mu
_{+}^{3}}-\frac{d^{3}I_{0}(\mu _{+})}{d\mu _{+}^{3}}\right] \right\}
(r-r_{+})  \nonumber \\
&&+\frac{1}{\pi ^{2}r_{+}^{6}}\left\{ \left[ \frac{m^{6}r_{+}^{6}}{16}
+\left( \xi -\frac{1}{12}\right) \frac{m^{4}r_{+}^{4}}{4}+\left( \xi -\frac{1
}{6}\right) \frac{\xi m^{2}r_{+}^{2}}{4}\right] \frac{d^{2}I_{0}(\mu _{+})}{
d\mu _{+}^{2}}\right.   \nonumber \\
&&+\frac{1}{\mu _{+}}\left[ -\frac{m^{6}r_{+}^{6}}{4}+\left( -\frac{\xi }{4}+
\frac{7}{192}\right) m^{4}r_{+}^{4}+\left( 2\xi ^{2}-\frac{59\xi }{96}+\frac{
3}{64}\right) m^{2}r_{+}^{2}+3\xi ^{3}\right.   \nonumber \\
&&\left. -\frac{11\xi ^{2}}{8}+\frac{5\xi }{24}-\frac{1}{96}\right] \frac{
dI_{0}(\mu _{+})}{d\mu _{+}}+\left( -4\xi +\frac{9}{16}\right) \mu
_{+}^{2}I_{1}(\mu _{+})-\frac{5}{4}\mu _{+}^{4}I_{2}(\mu _{+})  \nonumber \\
&&+\left[ \left( \xi -\frac{1}{6}\right) \frac{5m^{2}r_{+}^{2}}{16}+\frac{
5\xi ^{2}}{8}-\frac{5\xi }{24}+\frac{1}{48}\right] \ln \left( \frac{\mu
_{+}^{2}}{m_{\mbox{\tiny \sl DS}}^{2}r_{+}^{2}}\right) +\frac{1}{\mu _{+}^{2}
}\left[ \frac{7m^{4}r_{+}^{4}}{384}\right.   \nonumber \\
&&\left. \left. +\left( -\frac{5\xi ^{2}}{8}+\frac{3\xi }{16}-\frac{391}{
46080}\right) m^{2}r_{+}^{2}-\frac{5\xi ^{3}}{4}+\frac{11\xi ^{2}}{24}-\frac{
1111\xi }{23040}+\frac{241}{184320}\right] \right\} (r-r_{+})^{2}  \nonumber
\\
&&+\frac{1}{\pi ^{2}r_{+}^{7}}\left\{ \frac{1}{\mu _{+}}\left[ \frac{
m^{8}r_{+}^{8}}{48}+\left( \xi +\frac{1}{36}\right) \frac{m^{6}r_{+}^{6}}{8}
+\left( \frac{\xi ^{2}}{4}+\frac{\xi }{72}-\frac{1}{180}\right)
m^{4}r_{+}^{4}\right. \right.   \nonumber \\
&&\left. +\left( \frac{\xi ^{3}}{6}+\frac{\xi ^{2}}{72}-\frac{7\xi }{90}+
\frac{11}{720}\right) m^{2}r_{+}^{2}\right] \frac{d^{3}I_{0}(\mu _{+})}{d\mu
_{+}^{3}}+\frac{1}{\mu _{+}^{2}}\left[ -\frac{m^{8}r_{+}^{8}}{12}\right.
\nonumber \\
&&\left. +\left( \frac{\xi }{3}-\frac{35}{576}\right) m^{6}r_{+}^{6}+\left(
4\xi ^{2}-\frac{47\xi }{48}+\frac{17}{320}\right) m^{4}r_{+}^{4}+\left(
\frac{28\xi ^{3}}{3}-\frac{281\xi ^{2}}{48}\right. \right.   \nonumber \\
&&\left. \left. +\frac{1687\xi }{1440}-\frac{19}{288}\right) m^{2}r_{+}^{2}+
\frac{20\xi ^{4}}{3}-\frac{149\xi ^{3}}{18}+\frac{211\xi ^{2}}{72}-\frac{
71\xi }{180}+\frac{13}{720}\right] \frac{d^{2}I_{0}(\mu _{+})}{d\mu _{+}^{2}}
\nonumber \\
&&+\frac{1}{\mu _{+}^{3}}\left[ \frac{m^{8}r_{+}^{8}}{2}+\left( -\frac{5\xi
}{12}+\frac{23}{192}\right) m^{6}r_{+}^{6}+\left( -\frac{29\xi ^{2}}{2}+%
\frac{677\xi }{144}-\frac{2173}{5760}\right) m^{4}r_{+}^{4}\right.
\nonumber \\
&&\left. +\left( -37\xi ^{3}+\frac{2117\xi ^{2}}{144}-\frac{2447\xi }{1440}+
\frac{1}{20}\right) m^{2}r_{+}^{2}-\frac{82\xi ^{4}}{3}+\frac{104\xi ^{3}}{9}
-\frac{217\xi ^{2}}{160}\right.   \nonumber \\
&&\left. +\frac{19\xi }{1440}+\frac{7}{1920}\right] \frac{dI_{0}(\mu _{+})}{
d\mu _{+}}+\left( \frac{27\xi }{2}-\frac{49}{24}\right) \mu
_{+}^{2}I_{1}(\mu _{+})-\frac{5\mu _{+}^{4}}{2}I_{2}(\mu _{+})  \nonumber \\
&&+\left[ -\left( \xi -\frac{1}{6}\right) \frac{5m^{2}r_{+}^{2}}{8}-\frac{
5\xi ^{2}}{4}+\frac{5\xi }{12}-\frac{2}{45}\right] \ln \left( \frac{\mu
_{+}^{2}}{m_{\mbox{\tiny \sl DS}}^{2}r_{+}^{2}}\right) +\frac{1}{\mu _{+}^{4}
}\left[ -\frac{m^{6}r_{+}^{6}}{24}\right.   \nonumber \\
&&\left. +\left( \frac{5\xi ^{2}}{4}-\frac{41\xi }{144}-\frac{229}{23040}
\right) m^{4}r_{+}^{4}+\left( 5\xi ^{3}+\frac{101\xi ^{2}}{72}-\frac{3059\xi
}{2880}+\frac{323}{2880}\right) m^{2}r_{+}^{2}\right.   \nonumber \\
&&\left. \left. +5\xi ^{4}+\frac{65\xi ^{3}}{18}-\frac{13861\xi ^{2}}{5760}+
\frac{1001\xi }{2560}-\frac{7207}{368640}\right] \right\}
(r-r_{+})^{3}+O\left( (r-r_{+})^{4}\right) ,
\end{eqnarray}%
where
\begin{equation}
I_{n}(\mu _{+})=\int\nolimits_{0}^{\infty }\frac{x^{2n-1}\ln |1-x^{2}|}{
1+e^{2\pi |\mu _{+}|x}}dx,
\end{equation}
$m_{\mbox{\tiny \sl DS}}$ is equal to the mass $m$ of the field for a
massive scalar field. For a massless scalar field it is an arbitrary
parameter due to the infrared cutoff in renormalization counterterms for $
\left\langle T_{\nu }^{\mu }\right\rangle $. A particular choice of the
value of $m_{\mbox{\tiny \sl DS}}$ corresponds to a finite renormalization
of the coefficients of terms in the gravitational Lagrangian and must be
fixed by experiment or observation.

By direct calculations, one can check that the stress-energy tensor listed
above obeys the conservation law $\left\langle T_{\nu }^{\mu }\right\rangle
_{;\mu }=0$.

In the case $m=0,\ \xi =1/6$ we can numerically evaluate $I_{n}(\mu _{+})$
and the derivatives of these functions as follows
\begin{eqnarray}
I_{1}(\sqrt{1/12}) &\approx &-0.05962,\quad I_{2}(\sqrt{1/12})\approx
0.50384,  \nonumber \\
\quad \frac{dI_{0}(\mu _{+})}{d\mu _{+}} &\approx &0.64950,\quad \frac{
dI_{1}(\mu _{+})}{d\mu _{+}}\approx -0.66948,\quad \frac{dI_{2}(\mu _{+})}{
d\mu _{+}}\approx -11.70126,  \nonumber \\
\quad \frac{d^{2}I_{0}(\mu _{+})}{d\mu _{+}^{2}} &\approx &-0.03288,\quad
\frac{d^{2}I_{1}(\mu _{+})}{d\mu _{+}^{2}}\approx 21.17445,  \nonumber \\
\quad \frac{d^{3}I_{0}(\mu _{+})}{d\mu _{+}^{3}} &\approx &-43.75175,\quad
\frac{d^{3}I_{1}(\mu _{+})}{d\mu _{+}^{2}}\approx -462.08899
\end{eqnarray}%
and the expressions (\ref{1tt},\ref{eps}) take the form
\begin{eqnarray}  \label{00m0}
\langle T_{t}^{t}\rangle _{ren} &\simeq &\frac{1}{\pi ^{2}r_{+}^{4}}\left[
0.00078+\frac{1}{2880}\ln \left( m_{\mbox{\tiny \sl DS}}^{2}r_{+}^{2}\right)
\right] +\frac{1}{\pi ^{2}r_{+}^{5}}\left[ -0.00241\right.  \nonumber \\
&&\left. -\frac{1}{720}\ln \left( m_{\mbox{\tiny \sl DS}}^{2}r_{+}^{2}
\right) \right] \left( r-r_{+}\right) +\frac{1}{\pi ^{2}r_{+}^{6}}\left[
0.00463+\frac{1}{288}\ln \left( m_{\mbox{\tiny \sl DS}}^{2}r_{+}^{2}\right)
\right] \left( r-r_{+}\right) ^{2}  \nonumber \\
&&+\frac{1}{\pi ^{2}r_{+}^{7}}\left[ 0.00417-\frac{1}{90}\ln \left( m_{
\mbox{\tiny \sl DS}}^{2}r_{+}^{2}\right) \right] \left( r-r_{+}\right)
^{3}+O\left( \frac{\left( r-r_{+}\right) ^{4}}{r_{+}^{8}}\right) ,
\end{eqnarray}
\begin{equation}
\langle T_{r}^{r}\rangle _{ren}-\langle T_{t}^{t}\rangle _{ren}\simeq \frac{
1 }{\pi ^{2}r_{+}^{7}}\left[ -0.007406+\frac{1}{360}\ln \left( m_{
\mbox{\tiny \sl DS}}^{2}r_{+}^{2}\right) \right] \left( r-r_{+}\right) ^{3}
+O\left( \frac{\left( r-r_{+}\right) ^{4}}{r_{+}^{8}}\right) ,
\end{equation}
\begin{eqnarray}
\langle T_{\theta }^{\theta }\rangle _{ren} &\simeq &\frac{1}{\pi
^{2}r_{+}^{4}}\left[ -0.00043-\frac{1}{2880}\ln \left( m_{\mbox{\tiny \sl DS}
}^{2}r_{+}^{2}\right) \right] +\frac{1}{\pi ^{2}r_{+}^{5}}\left[
0.00102\right.  \nonumber \\
&&\left. +\frac{1}{720}\ln \left( m_{\mbox{\tiny \sl DS}}^{2}r_{+}^{2}
\right) \right] \left( r-r_{+}\right) +\frac{1}{\pi ^{2}r_{+}^{6}}\left[
-0.00115-\frac{1}{288}\ln \left( m_{\mbox{\tiny \sl DS}}^{2}r_{+}^{2}\right)
\right] \left( r-r_{+}\right) ^{2}  \nonumber \\
&&+\frac{1}{\pi ^{2}r_{+}^{7}}\left[ -0.06380+\frac{7}{720}\ln \left( m_{
\mbox{\tiny \sl DS}}^{2}r_{+}^{2}\right) \right] \left( r-r_{+}\right)
^{3}+O\left( \frac{\left( r-r_{+}\right) ^{4}}{r_{+}^{8}}\right) .
\end{eqnarray}

In the large mass field limit ($m_{\mbox{\tiny \sl DS}}^{2}=m^{2}\gg 2\xi
/r_{+}^{2}$) we obtain
\begin{equation}
\ln {\frac{\mu ^{2}}{m^{2}r_{+}^{2}}}=\frac{8\xi -1}{4m^{2}r_{+}^{2}}-\frac{
(8\xi -1)^{2}}{32m^{4}r_{+}^{4}}+\frac{(8\xi -1)^{3}}{192m^{6}r_{+}^{6}}
+O\left( \frac{1}{m^{8}r_{+}^{8}}\right) ,
\end{equation}
\begin{eqnarray}
I_{0}(\mu _{+}) &=&-\frac{1}{48\mu _{+}^{2}}-\frac{7}{3840\mu _{+}^{4}}-
\frac{31}{48384\mu _{+}^{6}}+O\left( \frac{1}{\mu _{+}^{8}}\right) =-\frac{1
}{48\;m^{2}r_{+}^{2}}+\frac{1}{m^{4}r_{+}^{4}}\left( \frac{\xi }{24}\right.
\nonumber \\
&&\left. -\frac{9}{1280}\right) +\frac{1}{m^{6}r_{+}^{6}}\left( -\frac{\xi
^{2}}{12}+\frac{9\xi }{320}-\frac{1381}{483840}\right) +O\left( \frac{1}{
m^{8}r_{+}^{8}}\right) ,  \nonumber \\
I_{1}(\mu _{+}) &=&-\frac{7}{1920\mu _{+}^{4}}-\frac{31}{32256\mu _{+}^{6}}
+O\left( \frac{1}{\mu _{+}^{8}}\right)   \nonumber \\
&=&-\frac{7}{1920\;m^{4}r_{+}^{4}}+\frac{1}{m^{6}r_{+}^{6}}\left( \frac{7\xi
}{480}-\frac{449}{161280}\right) +O\left( \frac{1}{m^{8}r_{+}^{8}}\right) ,
\nonumber \\
I_{2}(\mu _{+}) &=&-\frac{31}{16128\;\mu _{+}^{6}}+O\left( \frac{1}{\mu
_{+}^{8}}\right) =-\frac{31}{16128\;m^{6}r_{+}^{6}}+O\left( \frac{1}{
m^{8}r_{+}^{8}}\right)
\end{eqnarray}
and the expressions (\ref{1tt}), (\ref{eps}) coincide with the correspondent
expressions of the DeWitt-Schwinger approximation \cite{AHS,Mat1,Mat2}
\begin{eqnarray}
\langle T_{t}^{t}\rangle _{ren} &=&\frac{1}{\pi ^{2}m^{2}r_{+}^{6}}\left( -
\frac{\xi ^{3}}{24}+\frac{\xi ^{2}}{48}-\frac{\xi }{240}+\frac{1}{2520}
\right) +\frac{1}{\pi ^{2}m^{2}r_{+}^{7}}\left( \frac{\xi ^{3}}{4}-\frac{\xi
^{2}}{8}+\frac{\xi }{40}\right.   \nonumber \\
&&\left. -\frac{1}{420}\right) \left( r-r_{+}\right) +\frac{1}{\pi
^{2}m^{2}r_{+}^{8}}\left( -\frac{5\xi ^{3}}{8}+\frac{5\xi ^{2}}{16}-\frac{
47\xi }{720}+\frac{1}{144}\right) \left( r-r_{+}\right) ^{2}  \nonumber \\
&&+\frac{1}{\pi ^{2}m^{2}r_{+}^{9}}\left( \frac{5\xi ^{3}}{4}-\frac{5\xi ^{2}
}{8}+\frac{73\xi }{540}-\frac{221}{15120}\right) \left( r-r_{+}\right)
^{3}+O\left( \left( r-r_{+}\right) ^{4}\right) ,
\end{eqnarray}
\begin{eqnarray}
\langle T_{r}^{r}\rangle _{ren}-\langle T_{t}^{t}\rangle _{ren} &=&\frac{1}{
\pi ^{2}m^{2}r_{+}^{9}}\left( \frac{\xi }{270}-\frac{2}{945}\right) \left(
r-r_{+}\right) ^{3}  \nonumber \\
&&+O\left( \frac{(r-r_{+})^{3}}{m^{4}r_{+}^{11}}\right) +O\left(
(r-r_{+})^{4}\right) ,
\end{eqnarray}%
\begin{eqnarray}
\langle T_{\theta }^{\theta }\rangle _{ren} &=&\langle T_{\varphi }^{\varphi
}\rangle _{ren}=\frac{1}{\pi ^{2}m^{2}r_{+}^{6}}\left( \frac{\xi ^{3}}{12}-
\frac{\xi ^{2}}{24}+\frac{\xi }{120}-\frac{1}{1260}\right) +\frac{1}{\pi
^{2}m^{2}r_{+}^{7}}\left( -\frac{\xi ^{3}}{4}+\frac{\xi ^{2}}{8}\right.
\nonumber \\
&&\left. -\frac{\xi }{36}+\frac{17}{5040}\right) \left( r-r_{+}\right) +
\frac{1}{\pi ^{2}m^{2}r_{+}^{8}}\left( \frac{5\xi ^{3}}{8}-\frac{5\xi ^{2}}{
16}+\frac{29\xi }{360}-\frac{17}{1440}\right) \left( r-r_{+}\right) ^{2}
\nonumber \\
&&+\frac{1}{\pi ^{2}m^{2}r_{+}^{9}}\left( -\frac{5\xi ^{3}}{4}+\frac{79\xi
^{2}}{24}-\frac{35\xi }{27}+\frac{13}{90}\right) \left( r-r_{+}\right)
^{3}+O\left( \left( r-r_{+}\right) ^{4}\right) .
\end{eqnarray}



\section*{Acknowledgments}

We would like to thank S. V. Sushkov, N. R. Khusnutdinov, and J. Matyjasek
for helpful discussions. This work was supported in part by grants
05-02-17344, 05-02-39023 and 06-01-00765 from the Russian Foundation for
Basic Research.


\end{document}